\documentclass[pra,twocolumn,showpacs,floatfix]{revtex4-1}
\usepackage{amsmath}
\usepackage{bm}
\usepackage{graphicx}
\usepackage{amssymb}
\usepackage{datetime}
\usepackage{braket}
\usepackage{mathtools}
\usepackage{xcolor}
\usepackage[multidots]{grffile}
\usepackage{rotating}
\usepackage{array}
\usepackage{multirow}
\usepackage{siunitx}
\usepackage{times}
\begin{document}
\newcommand{\figdir}{.}
\newcommand{\figwidth}{0.95\columnwidth}
\newcommand{\ffigwidth}{0.4\columnwidth}
\newcommand{\un}{\mathcal{U}}
\newcommand{\cs}{\frac{\rm d\sigma}{\rm d\Omega}}
\newcommand{\cstext}{\rm d\sigma / \rm d\Omega}
\newcommand{\csel}{\left.\frac{\rm d\sigma}{\rm d\Omega}\right|_{\rm el}}
\newcommand{\csin}{\left.\frac{\rm d\sigma}{\rm d\Omega}\right|_{\rm inel}}
\newcommand{\bcsin}{\left.\frac{\rm d\sigma}{\rm d\Omega}\right|_{\rm inel}^{\rm Bog}}
\newcommand{\ecsin}{\left.\frac{\rm d\sigma}{\rm d\Omega}\right|_{\rm inel}^{\rm Exact}}
\newcommand{\qtil}{\tilde{q}}
\newcommand{\PreserveBackslash}[1]{\let\temp=\\#1\let\\=\temp}
\let\PBS=\PreserveBackslash
\title{Matter-wave scattering from interacting bosons in an optical lattice}
\newcommand{\freiburg}{Physikalisches Institut, Albert-Ludwigs Universit\"{a}t Freiburg, Hermann-Herder Stra{\ss}e 3, D-79104, Freiburg, Germany}
\author{Klaus Mayer}
\author{Alberto Rodriguez}
\author{Andreas Buchleitner}
\email[]{abu@uni-freiburg.de}
\affiliation{\freiburg}
\date{$Rev: 189 $, compiled \today, \currenttime}
%
\begin{abstract}
We study the scattering of matter-waves from interacting bosons in a one-dimensional optical lattice, described by the Bose-Hubbard Hamiltonian. We derive analytically a formula for the inelastic cross section as a function of the atomic interaction in the lattice, employing Bogoliubov's formalism for small condensate depletion. 
A linear decay of the inelastic cross section for weak interaction, independent of number of particles, condensate depletion and system size, is found. 
\end{abstract}

\maketitle

\section{Introduction}
\label{sec-introduction}
Scattering experiments have a long-standing tradition in most physical disciplines. After the probe has interacted with the target, it carries information on the latter, unraveling, e.g., the structure of the atom in Rutherford's experiments \cite{Rut1911}, or the make-up of nuclei in modern particle accelerators. Condensed-matter physics in particular has benefited from x-ray, electron and neutron scattering, which are well-established techniques to shed light on typical solid-state structures like crystals or amorphous materials. 

The realization of Bose-Einstein Condensation (BEC) \cite{DavMA95, AndEM95} 
and the subsequent development of 
optical traps \cite{WinS13}, has led to the implementation of Hubbard-like systems with an unprecedented experimental control, which made the investigation of 
many-body and strongly correlated physics feasible \cite{LewSAD07,BloDZ08,BloDN12}.    
Outstanding examples are the realization of the Mott Insulator to Superfluid quantum phase transition of bosons \cite{GreME02}, 
or the study of the interplay of disorder and interactions \cite{FalLG07,Mod10,PasMW10,DeiZR10,SanL10,DeiLM11}.
Since the high degree of experimental control 
provides access to the many-particle dynamics,
beyond a mere effective single-particle picture, the necessity arises to develop 
detection methods to obtain the relevant information. 
Considering the kinship of such experiments with 
solid-state systems, scattering techniques 
appear as a natural candidate \cite{YeZL13}.

Elastic and inelastic scattering of photons has been exploited for the analysis of ultracold gases, both theoretically 
\cite{ReyBP05,LakIT09,DouB11,DouB11b,YeZL11,SykB11,JacI12,RouMR13} and experimentally \cite{OzeKS05,DuWY10,CleFF09,ErnGK10,BisGL11,FabHC12}. Most recently, 
inelastic scattering of matter waves has been shown to allow a clear distinction of the Mott and superfluid 
states of bosons in an optical lattice \cite{SanMH10}. 
Experimentally, a cloud of Bose-condensed atoms has been used for Bragg-scattering off a second cloud trapped in an optical lattice, presenting matter-wave scattering as a suitable method for the characterization of strongly correlated phases of quantum gases \cite{GadPR12}. 
Additionally, the scattering of atoms 
can monitor the system in a non-destructive manner, and the influence of probe-induced excitations in the target on subsequent scattering events 
can be controlled \cite{DouB13}.

Here, we study the inelastic scattering of a matter-wave from a system of interacting bosons in a one-dimensional optical lattice. In particular, we present a thorough analytical and numerical analysis of the decay of the inelastic cross section as the interaction among the target bosons is increased. 
In Sec.~\ref{sec:scatt} 
we spell out the set problem and introduce the many-body cross section, as well as its limits for non-interacting and strongly-interacting bosons \cite{SanMH10}. 
In Sec.~\ref{sec:BogoAp}, we expand on Bogoliubov's approximation for a weakly-depleted BEC \cite{Bog47}, which we employ in Sec.~\ref{sec:CSbog} to derive analytically the inelastic cross section as a function of the interaction between two bosons. The analysis, discussion and comparison with numerical results  
are given in Sec.~\ref{sec:analysis}. 
Some technical details of the calculations are collected in the Appendix at the end. 

%
\section{Many-body scattering cross section}
\label{sec:scatt}
%
We study scattering of neutral atoms from a target comprised of interacting bosons 
suspended in a one-dimensional optical lattice potential. The target is commonly described by the discrete version of the Bose-Hubbard Hamiltonian \cite{JakBCGZ98,LewSAD07}, taking into account only the first band of the lattice:
\begin{equation}
 H_{\rm BH} = -J\sum_{\langle j, j'\rangle}\hat{c}_j^\dagger \hat{c}_{j'} + \sum_{j=1}^{L}\left\{\frac{U}{2}\hat{c}_j^\dagger \hat{c}_j^\dagger \hat{c}_j \hat{c}_j -\mu \hat{c}_j^\dagger \hat{c}_j\right\},
\label{eq:HBH}
\end{equation}
in terms of bosonic creation and annihilation operators $\hat{c}_j^\dagger, \hat{c}_j,$ on lattice site $j$,
\begin{equation}
	[\hat{c}_j, \hat{c}_{j'}^\dagger] = \delta_{jj'},
\label{eq:commrel}
\end{equation}
and where $J>0$ is the nearest-neighbour tunneling strength and we limit our considerations to $U>0$, with $U$ the energy of the repulsive binary on-site interaction.
The tunneling strength $J$ depends on the depth of the optical lattice, while the interaction $U$ is controlled by scattering lengths which can be modified using Feshbach resonances \cite{TimTHK99}.
The hopping term in $H_{\rm BH}$ includes all pairs $\langle j, j'\rangle$ of neighbouring lattice sites 
in a system with $L$ sites. 
We work in the grand-canonical ensemble, where the mean number of atoms $N$ is determined by the chemical potential $\mu$. 
The basic unit of length in the system is given by the lattice constant $d=\pi/k_L$, where $k_L$ is the wavenumber of the laser providing the lattice potential of depth $V_0$.
Our reference energy will be the recoil energy $E_r=\hbar^2 k_L^2/2 M$, where $M$ is the mass of the atoms in the lattice. 

\begin{figure}
 \centering 
 \includegraphics[width=\figwidth]{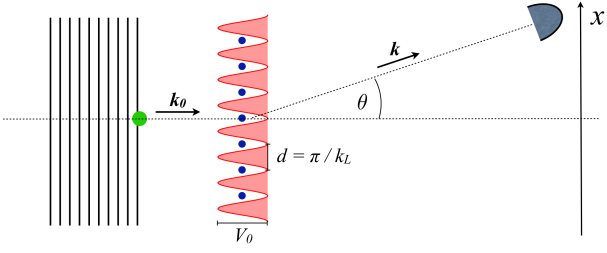}
\caption{(color online) Scattering setup: a particle of mass $m$ initially in a plane-wave state with momentum $\bm k_0$ is scattered into the angle $\theta$ from a target of atoms (all of which have mass $M$) submerged in a one-dimensional optical lattice with depth $V_0$ and lattice constant $d = \pi/k_L$, where $k_L$ is the laser wavenumber. The asymptotic final state of the probe has momentum $\bm k$.}
\label{fig:setup}
\end{figure}

We assume that the optical lattice is transparent to the scattering atom (probe), of mass $m$, and whose energy is not large enough to induce interband excitations in the target. 
For this low-energy probe the interaction with each atomic target will be dominated by $s$-wave scattering, and can be described by the pseudopotential \cite{Wod91,FetW03} 
\begin{equation}
 V(\bm r)= \frac{2\pi\hbar^2}{m} a_s \sum_{\beta=1}^N \delta(\bm r-\bm r^{(\beta)}), 
\end{equation}
with scattering length $a_s$, and where $\bm r$ and $\{\bm r^{(\beta)}\}_{\beta=1,\ldots,N}$ give the positions of the probe atom and the atoms in the lattice, respectively.
The initial state of the probe is assumed to be a plane wave of momentum ${\bm k_0}$ ($|{\bm k_0}| = k_0$), which gets scattered into an asymptotic final state with momentum ${\bm k}$ (cf.~Fig.~\ref{fig:setup}).
If the target is prepared in its ground state $\ket{\rm g}$ with corresponding energy $E_{\rm g}$, the many-body scattering cross section $\cstext$ in Born approximation, 
for a probe of initial energy $E_{0} = \hbar^2k_0^2/2m$, reads \cite{SanMH10,San10}:
\begin{align}
	\frac{1}{a_s^2}\cs =& \left|\int d{\bm r}\, e^{i{\bm\kappa \bm r}}\bra{\rm g}\hat{n}(\bm r)\ket{\rm g}\right|^2 \notag\\
	& + \sum_{\rm e}\sqrt{1-\frac{E_{\rm e}-E_{\rm g}}{E_{0}}}\left|\int d{\bm r}\, e^{i{\bm\kappa \bm r}}\bra{\rm e}\hat{n}(\bm r)\ket{\rm g}\right|^2,
\label{eq:crosssec_gen}
\end{align}
where $E_{\rm e}$ denotes the excitation spectrum  of $H_{\rm BH}$, with corresponding eigenstates $\ket{\rm e}$, and 
\begin{equation}
 \bm\kappa\equiv\bm k_0-\bm k
\end{equation}
is the transferred momentum, whose component in the direction $\textcolor{red}{\bm u_x}$ of the lattice, 
\begin{equation}
\kappa \equiv \bm \kappa \bm u_x,
\end{equation}
obeys 
\begin{equation}
 \kappa d = -\pi \sin\theta \sqrt{\frac{m}{M}\frac{E_0}{E_r}}\sqrt{1-\frac{E_{\rm e}-E_\textrm{g}}{E_0}},
 \label{eq:kappax}
\end{equation}
as follows from energy conservation. 
The first term in Eq.~\eqref{eq:crosssec_gen} corresponds to the elastic part of the cross section, whereas the sum in the second term runs over all excited states $\ket{\rm e}$ that are energetically allowed (i.e., for which $E_{\rm e}-E_{\rm g}<E_0$), whose contributions represent inelastic scattering.
The density operator, 
\begin{equation}
 \hat{n}(\bm r)=\hat{\Psi}^\dagger(\bm r)\hat{\Psi}(\bm r),
 \label{eq:densop}
\end{equation}
is defined by the bosonic field operators $\hat{\Psi}(\bm r)$, which can be expanded in the lattice basis:
\begin{equation}
	\hat{\Psi}(\bm r)=\sum_{j=1}^{L}\hat{c}_j w(\bm r - \bm r_j),
	\label{eq:psiexp}
\end{equation}
in terms of the Wannier functions $w(\bm r - \bm r_j)$ \cite{Koh59}, describing a particle localized at lattice site $j$, which in our one-dimensional system corresponds to $\bm r_j = x_j \bm u_x =jd\bm u_x$. 
The density operator can then be split into two contributions: diagonal and off-diagonal in the Wannier basis. 

Expression \eqref{eq:crosssec_gen} is valid in the far field to first order in the scattering potential.
The elastic part of the scattering cross section contains the single-particle Bragg-scattering signal ---represented by the Fourier transform of the atomic density in the first term of Eq.~\eqref{eq:crosssec_gen}--- resulting from the plane-wave initial state of the probe, and it is independent of the interaction $U$. In contrast, the inelastic cross section bears a clear signature of the interactions among the target atoms.
In the limits of vanishing and infinite $U/J$, analytical expressions for the cross section can be obtained \cite{SanMH10,San10}. 
These are summarized in Table \ref{tab:CS_MISF}, where the terms are grouped in contributions that are either diagonal or off-diagonal in the Wannier basis. The diagonal contributions are proportional to the form factor of a unit cell of the lattice,
\begin{align}
	W(\bm \kappa)=\int e^{i\bm\kappa\bm r}\left|w(\bm r)\right|^2d\bm r.
	\label{eq:Wofkappa}
\end{align}
On the other hand, off-diagonal terms are proportional to the overlap of two Wannier functions centered at different sites
\begin{align}
	W_{jl}(\bm\kappa) = \int e^{i\bm\kappa\bm r} w^*(\bm r - \bm r_j) w(\bm r - \bm r_l) d\bm r.
	\label{eq:Wnm}
\end{align}	
Within the validity of the approximations made in the derivation of the Bose-Hubbard Hamiltonian, one has that $|W_{j,j\pm1}(\bm\kappa)|/|W(\bm\kappa)|\sim 10^{-4}$, and the off-diagonal terms can be safely neglected for deep lattices. Throughout this paper, we consider a lattice depth of $V_0=15E_r$, which gives rise to a tunneling strength \mbox{$J=\SI{6.5e-3}{}E_r$}, with an energy gap to the second band of approximately $6 E_r$ \cite{Note1}.

\begin{table*}
\setlength\extrarowheight{3pt}
\begin{tabular}{c >{\centering}m{17mm}c >{\centering}m{10mm} c}
	\hline\hline 
	 & & { Diagonal}& &{ Off-diagonal}\\
	\hline\\[-3mm]
	\multirow{2}{*}[-4mm]{\parbox{24mm}{$U/J\rightarrow 0$\\ (Superfluid)}} & Elastic &  $\displaystyle \frac{N^2}{L^2} \bigg|\sum_{j}e^{i\bm\kappa_{\rm el}\bm r_j}W(\bm\kappa_{\rm el})$ & + & 
	$\displaystyle \sum\limits_{j\neq l}W_{jl}(\bm\kappa_{\rm el})\bigg|^2$\\[4mm]
	& Inelastic & $\displaystyle\frac{N}{L^2}\sum_{\bm q\neq0}C_{\rm sf}(\bm q)\bigg|\sum\limits_{j}e^{i(\bm\kappa^{\rm sf}_{\bm q}-\bm q)\bm r_j}W(\bm\kappa_q^{\rm sf})$ &+& 
	$\displaystyle\sum_{j\neq l}e^{-i\bm q\bm r_j} W_{jl}(\bm\kappa_{\bm q}^{\rm sf})\bigg|^2$ \\[8mm]
	\multirow{2}{*}[-3mm]{\parbox{24mm}{$U/J\rightarrow \infty$\\ (Mott insulator)}} & Elastic & $\displaystyle \frac{N^2}{L^2} \bigg|\sum_j e^{i\bm\kappa_{\rm el}\bm r_j}W(\bm\kappa_{\rm el})\bigg|^2$ &+& 0\\[4mm]
	&Inelastic & 0 &  + & $\displaystyle \frac{N}{L}\left(\frac{N}{L}+1\right)C_{\rm mi}\sum_{j\neq l}\left|W_{jl}(\bm\kappa_{\bm q}^{\rm mi})\right|^2$\\
	\hline\hline
\end{tabular}
\caption{Analytical expressions for the elastic and inelastic parts of the cross section, $a_s^{-2}\cstext$, in the limits ${U/J\rightarrow0}$ (superfluid) and ${U/J\rightarrow\infty}$ (Mott insulator). Different columns list the diagonal and off-diagonal terms with respect to the Wannier basis [see Eqs.~\eqref{eq:Wofkappa} and \eqref{eq:Wnm}]. 
The weighting factors read $C_{\rm sf}(\bm q) = \sqrt{1-[\epsilon(\bm q) - \epsilon(\bm 0)]/E_0}$ and \mbox{$C_{\rm mi} = \sqrt{1-U/E_0}$}, where $\epsilon(\bm q)$ is the single-particle Bloch dispersion relation; in the 1D case $\epsilon(\bm q) =  4J\sin^2{(qd/2)}$ where $q=2\pi s/(Ld)$ for $s=0,1,\ldots,L-1$. The $x$-component of the transferred momentum in the elastic case $\kappa_{\rm el}\equiv \bm\kappa_{\rm el} \bm u_x$ obeys  $\kappa_{\rm el} d=-\pi\sin\theta\sqrt{E_0m/E_rM}$, while in the inelastic case it fulfills $\kappa_q^{\rm sf} =\kappa_{\rm el} C_{\rm sf}(\bm q)$ and $\kappa_q^{\rm mi}=\kappa_{\rm el} C_{\rm mi}$, for the superfluid and Mott-insulating limits, respectively.}
\label{tab:CS_MISF}
\end{table*}

\begin{figure}
 \includegraphics[width=\figwidth]{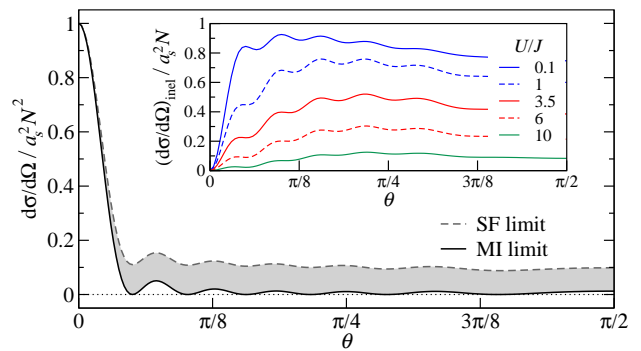}
\caption{(color online) Full scattering cross section normalized by $N^2$ in the Superfluid (SF) and Mott insulating (MI) limits, from the expressions given in Table  \ref{tab:CS_MISF} using the harmonic approximation for the Wannier functions [see Eq.~\eqref{eq:wann_gauss}], for $N=9$ particles in a lattice of length $L=9$. The scattering signal is symmetric around $\theta=0$. The incoming energy is $E_{0} = 2E_r$, the hopping energy is set to $J=0.0065 E_r$ and the masses of the probe and target atoms are taken to be equal. The elastic part of the cross section is the same in both regimes. The shaded region highlights the inelastic scattering component. 
The inset shows the inelastic cross section normalized by $N$, obtained from Eq.~\eqref{eq:crosssec_gen} and the numerically calculated spectrum of the system, for different values of the interaction $U$.}
\label{fig:CS_MISF}
\end{figure}

In the limit \mbox{$U/J\rightarrow0$}, all atoms condense in the same delocalized Bloch single-particle ground state of the optical lattice, and the system is found in a gapless superfluid (SF) phase. In the strongly interacting limit \mbox{$U/J\rightarrow\infty$}, and for integer filling factor $n=N/L$, the atoms localize on individual sites of the lattice (for $J=0$ the many-body ground state is given by a tensor product of Fock states), and the system is found in a gapped and incompressible Mott insulator (MI) phase \cite{FisWGF89,LewSAD07,EjiFGM12}.  
These two regimes give rise to markedly different inelastic scattering signals (Fig.~\ref{fig:CS_MISF}) \cite{SanMH10}. 
In the SF limit, the delocalized nature of the ground state results in a non-zero diagonal inelastic cross section, that scales with the number of atoms $N$, and is determined by the single particle Bloch dispersion relation of the lattice. 
On the other hand, in the MI limit the ground state becomes an eigenstate of the diagonal part of the density operator, and thus the diagonal contribution to the inelastic cross section is strictly zero (see Table \ref{tab:CS_MISF}). Due to the small overlap integral $W_{jl}(\bm\kappa)$, the off-diagonal excitations (i.e.~moving atoms from one site to another) are strongly suppressed, which leads to a vanishing inelastic scattering when approaching the MI limit, even when the probe energy is above the energy gap, $U$.
The elastic cross section scales as $N^2$ and, after neglecting the off-diagonal terms, is identical in both the SF and the MI limits. 

The transition between the SF and MI limits is then characterized by the decay of the inelastic cross section as a function of $U/J$ (Fig.~\ref{fig:CS_MISF}) \cite{SanMH10}. 
Here, we want to understand the way in which the interactions among the target atoms determine the emergence of this decay. 
To this end, we perform a Bogoliubov approximation of $H_{\rm BH}$ for small condensate depletion, and determine the dependence on $U$ of the many-body cross section, Eq.~(\ref{eq:crosssec_gen}).
%
\section{The Bogoliubov approximation for weak interaction}
\label{sec:BogoAp}
%
In the zero-temperature, ideal (i.e.~non-interacting) Bose gas, all $N$ bosons occupy the single-particle ground state of the Hamiltonian, thus called the condensate. The main idea underlying the Bogoliubov approach \cite{Bog47,NozPin:QLII} is that, for weak interaction, still most particles remain in this very state. Assuming that the ground state occupation $N_0$ is macroscopic, $N_0\gg1$, the non-commutativity of the creation and annihilation operators of a boson in the ground state, $\hat{b}_0^\dagger$ and $\hat{b}_0$, respectively, is neglected:
\begin{align}
	\langle\hat{b}_0^\dagger\hat{b}_0\rangle = N_0&\approx N_0+1 = \langle\hat{b}_0\hat{b}_0^\dagger\rangle\\
	\Rightarrow\left[\hat{b}_0,\hat{b}_0^\dagger\right]&\approx 0,
\label{eq:GSCommutator}
\end{align}
where the expectation values 
are to be taken with respect to the ground state. 
This implies that, physically, the \emph{state} of the system is not noticeably changed by removing (adding) a particle from (to) the condensate, and thus the replacement of both, $\hat{b}_0^\dagger$ and $\hat{b}_0$ with the 
value $\sqrt{N_0}$ is justified. 
This approach is termed non-number-conserving and constitutes a breaking of a $U(1)$-symmetry, since it assigns a fixed phase to the condensate:
\begin{align}
	\langle \hat{b}_0 \rangle = \langle \hat{b}_0^\dagger \rangle = \sqrt{N_0}.
	\label{eq:expvaluefield}
\end{align}

The effect of interactions among the bosons is to populate higher-energy single-particle states, which will be treated as small fluctuations.
In (discrete) position space, the field operator at position $x$, $\hat{c}_x$, is then written as:
\begin{align}
	\hat{c}_x = \phi_x + \delta \hat{c}_x,
\label{eq:Bog_gen}
\end{align}
where the {\it c}-number $\phi_x$ contains the ground-state contribution, and $\delta\hat{c}_x$ the fluctuation-operator contributions, respectively.
The formal replacement of 
Eq.~\eqref{eq:Bog_gen} in the Hamiltonian $H_{\rm BH}$ allows for the grouping of the different terms arising 
by their order in the fluctuation operators $\delta\hat{c}_x$:
\begin{equation}
	H_{\rm BH} = H_0 + H_1 + H_2 + H_3 + H_4.
	\label{eq:H1234}
\end{equation}
The actual approximation consists in neglecting terms of third and higher order, i.e.~$H_3$ and $H_4$. The validity of this truncation will be determined \emph{a posteriori} from the depletion of the condensate.
The equation of motion for the condensate wave function can be derived from the principle of least action, by considering independent variations of its real and imaginary parts \cite{PitS03,PetS08}. The time-independent description ensues by requiring that $H_0$ be stationary against small variations of $\phi_x^*$, so that the functional derivative of $H_0$ with respect to $\phi_x^*$ yields the discrete non-linear Schr\"odinger equation for $\phi_x$ \cite{Note2},
\begin{equation}
	\mu \phi_x = -J\sum_{\langle x,x'\rangle}\phi_{x'}+U|\phi_x|^2\phi_x,
\label{eq:DNLS}
\end{equation}
equivalently termed the Gross-Pitaevskii equation for the condensate mean field \cite{DalGPS99}. The chemical potential, $\mu$, can also be interpreted as a Lagrange multiplier fixing the mean number of particles in a variational derivation of Eq.~\eqref{eq:DNLS} \cite{Leg01}.
In a pure, one-dimensional periodic system, the single-particle eigenstates of the Hamiltonian \eqref{eq:HBH} are given by Bloch waves 
characterized by corresponding quasimomenta $q$, 
the ground state being the $q=0$ state. 
The value of $\phi_x$ can be read off by expanding $\hat{c}_x$ in momentum space:
\begin{align}
	\begin{split}
	\hat{c}_x &= \frac{1}{\sqrt{L}}\sum_{q\in {\rm BZ}} e^{iqx}\hat{b}_q\\
	 &= \sqrt{\frac{N_0}{L}} + \frac{1}{\sqrt{L}}\sum_{\substack{q\in {\rm BZ} \\ q\neq0}} e^{iqx}\hat{b}_q \label{eq:FourTrans}\\
	 &\equiv \sqrt{n_0} + \delta\hat{c}_x,\end{split}
\end{align}
where $n_0 = N_0/L$ denotes the (dimensionless) condensate density (or condensate filling factor), and BZ the first Brillouin zone of the lattice. For a lattice of length $L$, assuming periodic boundary conditions, the allowed values for the quasimomentum are \mbox{$q=2\pi s/(Ld)$} for \mbox{$s=0,1,\ldots,L-1$}.
Equation \eqref{eq:FourTrans} shows that the fluctuations are given by the occupation of higher momentum states, and \mbox{$\phi_x\equiv\sqrt{n_0}$}, for which Eq.~\eqref{eq:DNLS} yields the mean-field value of the chemical potential
\begin{equation}
	\mu = U n_0 -2J,
\label{eq:ChemPot}
\end{equation}
which is easily understood: adding one boson to the system requires the mean interaction energy with the present density of atoms and the (negative) tunneling energy to the two nearest neighbours.
We emphasize that in Eq.~\eqref{eq:H1234}, the order in the fluctuations is complementary to the leading order in which the amplitudes $\phi_x$ appear, i.e.~$H_0$ is proportional to $\phi_x^4$, $H_1$ to $\phi_x^3$, etc. Since the condensate amplitude is proportional to $\sqrt{N_0}$, the truncation of the Hamiltonian amounts to neglecting terms of order $\sqrt{N_0}$ or lower in the number of condensed particles. Therefore, quantities derived from the approximated Hamiltonian should only be given to that order of accuracy, as well. 

In one dimension, true Bose-Einstein condensation is forbidden, both at $T=0$ \cite{PitS91}, and for finite temperatures by the Mermin-Wagner-Hohenberg (MWH) theorem \cite{Hoh67,MerW66}, because the system displays no long-range off-diagonal order, i.e.~correlations in the single-particle-density matrix decay over a finite distance \cite{PitS03}. Instead, one has to work with the concept of a quasicondensate \cite{Pop72}, with a finite, but macroscopic correlation length. Keeping this in mind, throughout this work, we will still refer to the quasicondensate as \emph{condensate}. In Ref.~\cite{MorC03}, it has been shown that the Bogoliubov treatment of the one-dimensional problem requires a careful definition of the density and phase operators choosing a discretisation of space, which in our case is given naturally by the spacing $d$ of the optical lattice.
The validity of the treatment requires a large filling factor,
\begin{align}
	\frac{N}{L} =  n\gg1,
\end{align}
and the lattice spacing to be smaller than both, the thermal de-Broglie wavelength \mbox{${\lambda_{\rm dB} = \sqrt{2\pi\hbar^2/(Mk_{\rm B}T)}}$}, and the healing length \mbox{$\zeta = \sqrt{\hbar^2/(M|\mu|)}$}. Throughout this work we assume the zero-temperature limit, thus 
\mbox{${d<\lambda_{\rm dB}}$} is safely fulfilled. The remaining requirement,
\begin{equation}
    d<\zeta,\label{eq:req_phasdens_3}
\end{equation}
restricts the values of the interaction and tunneling energies.
From Eq.~(\ref{eq:ChemPot}), the latter condition translates into 
\begin{equation}
 2\left(1-\frac{E_r}{\pi^2 J}\right)< \frac{Un_0}{J} < 2\left(1+\frac{E_r}{\pi^2 J}\right),
 \label{eq:Jrestr}
\end{equation}
which for the tunneling strength \mbox{$J = \SI{6.5e-3}{\it E_r}$}, used in our calculations, yields the following restriction on the repulsive interaction energy
\begin{align}
	0<\frac{U n_0}{J}<33.
\label{eq:UrestrNumJ}
\end{align}
In the allowed regime, the procedure of Ref.~\cite{MorC03} leads to the same results as the usual Bogoliubov theory in density-phase representation \cite{GauM13}, which we discuss in the following. 

The many-body cross section \eqref{eq:crosssec_gen} involves matrix elements of the density operator $\hat{n}(\bm r)$. Within the Bogoliubov framework, this is treated most intuitively in the density-phase representation of the field operators, \cite{MorC03, GauM13}
\begin{equation}
	\hat{c}_x=e^{i\delta\hat{\varphi}_x}\sqrt{n_x + \delta\hat{n}_x},
\label{eq:cxdensphas}
\end{equation}
where $n_x$ denotes the mean density
at site $x$, and $\delta\hat{\varphi}_x$, $\delta\hat{n}_x$, are hermitian operators describing fluctuations of phase and density, respectively. 
A comparison of the first-order expansion in the fluctuations of Eq.~(\ref{eq:cxdensphas}) with Eq.~(\ref{eq:Bog_gen}) yields the identifications
\begin{align}
	\delta\hat{c}_x &= \frac{1}{2}\frac{\delta\hat{n}_x}{\sqrt{n_0}}+i\delta\hat{\varphi}_x\sqrt{n_0} \label{eq:delcx},\\
	\delta\hat{n}_x &= \sqrt{n_0}(\delta\hat{c}_x^\dagger + \delta\hat{c}_x) \label{eq:delnx},\\
	\delta\hat{\varphi}_x &= \frac{i}{2\sqrt{n_0}}(\delta\hat{c}_x^\dagger - \delta\hat{c}_x)\label{eq:delphi},
\end{align}
with $n_x\equiv n_0$.
From Eq.~\eqref{eq:FourTrans}, the density and phase fluctuation operators \eqref{eq:delnx} and \eqref{eq:delphi} can be cast as 
\begin{align}
		&\delta \hat{n}_x = \frac{1}{\sqrt{L}}\sum_{q\neq0}e^{iqx}\delta \hat{n}_q,\label{eq:dens_fluc_FT}\\
		&\delta \hat{\varphi}_x = \frac{1}{\sqrt{L}}\sum_{q\neq0}e^{iqx}\delta \hat{\varphi}_q, \label{eq:phas_fluc_FT}
\end{align}
where the hermiticity of $\delta\hat{n}_x$ ($\delta \hat{\varphi}_x$) requires \mbox{$\delta \hat{n}_q=\delta \hat{n}_{-q}^\dagger$} (\mbox{$\delta \hat{\varphi}_q = \delta \hat{\varphi}_{-q}^\dagger$}).
Since the condensate amplitude \mbox{$\phi_x=\sqrt{n_0}$} minimises $H_0$ via Eq.~\eqref{eq:ChemPot}, the Hamiltonian linear in the fluctuations $H_1$ vanishes, and 
the first non-vanishing correction to the mean field is given by the quadratic fluctuation Hamiltonian $H_2$~\cite{GauM13}, 
\begin{equation}
	H_2 = \sum_{q\neq0}\left\{\frac{1}{4n_0}(\epsilon_q+2Un_0)\delta\hat{n}_q^\dagger\delta\hat{n}_q + n_0 \epsilon_q\delta \hat{\varphi}_q^\dagger\delta \hat{\varphi}_q\right\},
\label{eq: H_2DP}
\end{equation}
with the dispersion of the lattice
\begin{equation}
	\epsilon_q = 4J\sin^2{(qd/2)}.
\label{eq:lattdisp}
\end{equation}
Note that in Eq.~(\ref{eq: H_2DP}) we have dropped constant terms 
(which can be accounted for by a redefinition of the energy origin), 
since we will only be interested in energy differences.
The quadratic Hamiltonian is diagonalised by a Bogoliubov transformation to a quasiparticle basis $\{\hat{\gamma}_q^\dagger,\hat{\gamma}_q\}_{q\neq0}$. It represents a canonical transformation, i.e.~the quasiparticle operators also obey bosonic commutation relations, Eq.~\eqref{eq:commrel}. This restriction, together with the diagonalisation requirement, leads to the explicit form of the transformation:
\begin{equation}
	\left(\begin{array}{c}
		\hat{\gamma}_q\\
		\hat{\gamma}_{-q}^\dagger		
	\end{array}\right) = \underline{A_q}\left(\begin{array}{c}
		i\sqrt{n_0}\,\delta\hat{\varphi}_q\\
		\frac{1}{2\sqrt{n_0}}\,\delta\hat{n}_q		
	\end{array}\right),\quad \underline{A_q} = \left(\begin{array}{c c} a_q & a_q^{-1} \\ -a_q & a_q^{-1}\end{array}\right),
\label{eq:QPTransf}
\end{equation}
where $a_q = \sqrt{\epsilon_q/\omega_q}$ and $\omega_q$ is the Bogoliubov dispersion,
\begin{equation}
	\omega_q = \sqrt{\epsilon_q(\epsilon_q + 2Un_0)}.
\label{eq:BogDisp}
\end{equation}
From Eq.~\eqref{eq:QPTransf} the Hamiltonian $H_2$ reduces to a collection of non-interacting quasiparticles with quasimomentum $q$ and energies $\omega_q$,
\begin{equation}
	H_2 = \sum_{q\neq0}\omega_q\hat{\gamma}_q^\dagger\hat{\gamma}_q
\label{eq: H_2QP}
\end{equation}
(again neglecting constants), whose ground state is the vacuum of quasiparticles, $\hat{\gamma}_q\ket{\rm g} = 0$, corresponding to the condensate. The eigenstates of this Hamiltonian are number states of the quasiparticle operators and represent fluctuations of density and phase of the condensate. 
We note that this is also true in a number-conserving approach \cite{CasD98}, which provides the same excitation spectrum. 
Since such approaches are somewhat more cumbersome, in this work we rely on the non-number-conserving formalism. 

Due to interaction-induced quantum fluctuations, even at zero temperature a non-zero depletion of the condensate is present, given by the density of particles with $q\neq0$ in the ground state:
\begin{align}
	\begin{split}
	n &= \frac{1}{L}\sum_{x}\langle \hat{c}_x^\dagger \hat{c}_x\rangle = n_0 + \frac{1}{L}\sum_{x}\langle\delta \hat{c}_x^\dagger\delta \hat{c}_x\rangle\\
	&= n_0 + \frac{1}{L}\sum_{q\neq0}\left\{\frac{\epsilon_q +U n_0}{2\omega_q}-\frac{1}{2}\right\},
\label{eq:N_depl}
	\end{split}
\end{align}
where the expectation values (which are to be taken with respect to the quasiparticle vacuum) of the fluctuations vanish, \mbox{$\langle\delta \hat{c}_x\rangle\equiv0$}, and the sum represents the depleted density, \mbox{${\delta n=n-n_0}$}.
Since the chemical potential $\mu$ ensures a fixed mean density $n$, Eq.~\eqref{eq:N_depl} can be numerically solved in order to obtain the relative depletion \mbox{$\delta n/n$}. It is this parameter that controls the validity of the truncation of the Hamiltonian \eqref{eq:H1234}: 
\mbox{$\delta n/n\ll1$} implies, that the number of bosons $N_0$ in the condensate is large, provided that the total number of bosons $N$ is large. 
In a one-dimensional lattice, for fixed $U$ and density $n$, the relative depletion goes to $1$ as the system size increases, and consequently the Bogoliubov approximation breaks down (in agreement with the absence of BEC in the thermodynamic limit in one dimension, according to MWH). On the other hand, for fixed $U$ and system size $L$, the relative depletion vanishes as $N^{-1/2}$ with the number of atoms for $N\gg1$. Thus, in 1D the quality of the Bogoliubov treatment always improves by increasing the number of bosons. 

\begin{figure}
 \includegraphics[width=\figwidth]{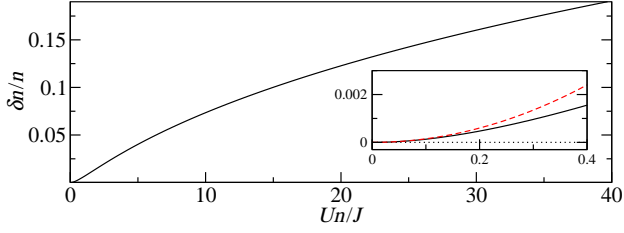}
\caption{(color online) Relative condensate depletion $\delta n/n$ as a function of the interaction parameter $\un=Un/J$, for $N=20$ particles on $L=5$ sites. The inset shows the behavior of $\delta n/n$ for small $\un$ (black, solid line) and the quadratic approximation (red, dashed), Eq.~\eqref{eq:depl_quadrappr}.}
\label{fig:depl_1}
\end{figure}

We introduce the dimensionless interaction parameter 
\begin{equation}
 \un\equiv\frac{Un}{J},
\end{equation}
which gives essentially the ratio of the interaction and the kinetic energies in the system. 
For a fixed system size $L$ and density $n$, the relative depletion grows monotonically with $\un$, as shown in Fig.~\ref{fig:depl_1}.
For integer filling factor, the condensate density has to vanish at the SF-MI phase transition, i.e.~for a finite value of $\un$. However, \mbox{$\delta n/n$}, as obtained from Eq.~\eqref{eq:N_depl}, approaches $1$ only for \mbox{$\un\rightarrow\infty$}, irrespective of $n$ being integer. Therefore, the SF-MI phase transition is not captured in the Bogoliubov-formalism \cite{OosSS01}.

In the regime $\un\ll1$, we can expand Eq.~\eqref{eq:N_depl} 
and obtain a quadratic dependence for the relative depletion (cf. Fig.~\ref{fig:depl_1}):
\begin{equation}
	\frac{\delta n}{n} = \alpha\, \un^2, 
\label{eq:depl_quadrappr}
\end{equation}
with \mbox{$\alpha=(L^4+10L^2-11)/(2880N)$}. The range of validity of this quadratic behavior decreases with the system size roughly as $L^{-2}$. Nevertheless, Eq.~\eqref{eq:depl_quadrappr} entails that   
as \mbox{$\un\rightarrow0$} the relative depletion goes to zero with a vanishing slope for any finite $L$. 
We will see below that this feature 
determines the decay 
of the inelastic cross section for weak interaction. 
%
\section{Expression for the cross section in Bogoliubov approximation}
\label{sec:CSbog}
%
The inelastic part of the many-particle scattering cross section, Eq.~\eqref{eq:crosssec_gen}, is comprised of a sum over all energetically allowed excited states $\ket{\rm e}$ 
with a non-vanishing matrix element 
$\bra{\rm e}\hat{n}(\bm r)\ket{\rm g}$ of the density operator $\hat{n}(\bm r)$. 
 As described in Sec.~\ref{sec:scatt}, the spatial representation of the density operator is conveniently obtained from the lowest-band Wannier basis of the lattice  $w(x-x_j)$ [see Eqs.~\eqref{eq:densop} and \eqref{eq:psiexp}], where $x_j=jd$ is the center of the $j$-th lattice site.
 Furthermore, only the diagonal elements of the density operator in this basis are considered. 
 In the Bogoliubov framework, we expand the latter elements in the fluctuations, using Eqs.~\eqref{eq:Bog_gen} and \eqref{eq:delcx}, and the inelastic cross section reads 
\begin{align}
		\frac{1}{a_s^2}\csin =& \sum_{\rm e}\sqrt{1-\frac{E_{\rm e}-E_{\rm g}}{E_{0}}}\left|W(\kappa_{\rm e})\right|^2 \times \notag\\ 
		&\left|\sum_{j = 1}^{L} e^{i\kappa_{\rm e} x_j}\bra{\rm e}\delta \hat{n}_j+\frac{1}{4n_0}\delta \hat{n}_j^2 + n_0\delta \hat{\varphi}_j^2\ket{\rm g}\right|^2,
\label{eq:BogCS_1}
\end{align}
where 
for simplicity we denote $\delta \hat{n}_{x_j}\equiv \delta \hat{n}_j$, $\delta \hat{\varphi}_{x_j}\equiv \delta \hat{\varphi}_j$, and the $x$-component of the transferred momentum is labeled as $\kappa_{\rm e}$ to emphasize its dependence on the excited state. Let us recall that the transferred momentum is given by Eq.~\eqref{eq:kappax}, and 
the form factor of the lattice unit cell, $W(\kappa)$, is defined in Eq.~\eqref{eq:Wofkappa}.

To linear order, the fluctuations in Eq.~\eqref{eq:BogCS_1} are density fluctuations, and phase fluctuations only appear to second order. 
By virtue of Eqs.~\eqref{eq:dens_fluc_FT}, \eqref{eq:phas_fluc_FT} and \eqref{eq:QPTransf} the fluctuations are transformed into the quasiparticle basis, yielding 
\begin{widetext}
\begin{align}
		\frac{1}{a_s^2}\csin =& \sum_{\rm e}\sqrt{1-\frac{E_{\rm e}-E_{\rm g}}{E_{0}}}\left|W(\kappa_{\rm e})\right|^2 \left|\sum_{j=1}^{L} e^{i\kappa_{\rm e} x_j}\left\{\frac{1}{\sqrt{L}}\sum_{q\neq0}e^{iqx_j}\sqrt{n_0}a_q\bra{\rm e}(\hat{\gamma}_q+\hat{\gamma}_{-q}^\dagger)\ket{\rm g} \right.\right. \notag\\ & \left.\left. + \frac{1}{4L}\sum_{q,q'\neq0}e^{i(q+q')x_j}\left[a_qa_{q'}\bra{\rm e}(\hat{\gamma}_q+\hat{\gamma}_{-q}^\dagger)(\hat{\gamma}_{q'}+\hat{\gamma}_{-q'}^\dagger)\ket{\rm g}-a_q^{-1}a_{q'}^{-1}\bra{\rm e}(\hat{\gamma}_q-\hat{\gamma}_{-q}^\dagger)(\hat{\gamma}_{q'}-\hat{\gamma}_{-q'}^\dagger)\ket{\rm g} \right]\right\}\right|^2.
\label{eq:BogCS_2}
\end{align}
\end{widetext}
Since the excited states are number states in the quasiparticle basis, the sum over $\ket{\rm e}$ in Eq.~\eqref{eq:BogCS_2} becomes a sum over quasiparticle modes and can be split into two contributions: one corresponding to the excitation of one quasi-particle, and a second one, corresponding to the simultaneous excitation of two quasiparticles:
\begin{widetext}
\begin{equation}
		\frac{1}{a_s^2}\csin = \frac{N_0}{L^2}\sum_{q\neq0}\sqrt{1-\frac{\omega_q}{E_{0}}}\frac{\epsilon_q}{\omega_q}\left|\Sigma(\kappa_q-q)W(\kappa_q)\right|^2
		+\frac{1}{2L^2}\sum_{q,q'\neq0}\sqrt{1-\frac{\omega_q+\omega_{q'}}{E_{0}}}f(q,q')\left|\Sigma\left(\kappa_{q+q'}-(q+q')\right)W(\kappa_{q+q'})\right|^2,
\label{eq:BogCS_3}
\end{equation}
\end{widetext}
where
\begin{align}
	\Sigma(\kappa_q-q)&=\sum_{j=1}^L e^{i(\kappa_q-q)x_j},
	\label{eq:sigma}\\ 
	|\Sigma(\kappa_q-q)|^2&=
	\frac{\sin^2\left((\kappa_q-q)dL/2\right)}{\sin^2\left((\kappa_q-q)d/2\right)},
\end{align}
and
\begin{equation}
	f(q,q') = \frac{\epsilon_q\epsilon_{q'}+Un_0(\epsilon_{q}+\epsilon_{q'}) + 2(Un_0)^2-\omega_q\omega_{q'}}{(1+\delta_{q,q'})\,\omega_q\omega_{q'}}.
\end{equation}
The $x$-component of the transferred momenta, \mbox{$\kappa_q$} and 
\mbox{$\kappa_{q+q'}$}, are now characterized by the quasimomenta of the excitations, and are evaluated via Eq.~\eqref{eq:kappax}, replacing the excitation energy $E_{\rm e}-E_{\rm g}$ by $\omega_q$ 
and \mbox{$\omega_q+\omega_{q'}$}, corresponding to the creation of one 
quasiparticle in mode $q$, and two quasiparticles in 
modes $q$ and $q'$, respectively.
In Eq.~\eqref{eq:BogCS_3}, we have stated the two-quasiparticle contribution for completeness. It should, however, be dropped: while the single-quasiparticle contribution scales like $N_0$, the two-quasiparticle contribution is of order one. 
Terms of this order have already been 
neglected in the truncation of Hamiltonian \eqref{eq:H1234}; they are therefore not complete and must be neglected here as well. 
Thus, the inelastic many-particle cross section in Bogoliubov approximation is given by the first term of Eq.~\eqref{eq:BogCS_3}. 
In order to compare different configurations, we normalize the cross section to the total number of bosons $N$, which is the scale of the superfluid cross section (cf.~Table \ref{tab:CS_MISF}), and obtain finally 
\begin{equation}
	\frac{1}{Na_s^2}\bcsin= \frac{1}{L^2}\sum_{q\neq0}\sqrt{1-\frac{\omega_q}{E_{0}}}\frac{n_0\,\epsilon_q}{n\,\omega_q}\left|\Sigma(\kappa_q-q)W(\kappa_q)\right|^2.
\label{eq:BogCSfinal}
\end{equation}
This latter expression corresponds to the term linear in the fluctuations in Eq.~\eqref{eq:BogCS_1}, which shows that to this level of the approximation the inelastic cross section stems from the excitation of density fluctuations only \cite{NozPin:QLII}. In fact, the quantity $\epsilon_q/\omega_q$ corresponds to the dynamic structure factor in Bogoliubov approximation, which characterizes the system's response to a density perturbation \cite{MenKPS03, RouMR13}.

The result \eqref{eq:BogCSfinal} allows for the analysis of the dependence of the inelastic cross section 
on the system parameters, in particular on the interaction strength $\un$.
For this task, the Wannier function will be approximated by a Gaussian, 
corresponding to the ground state of the harmonic approximation of each potential well of the lattice: 
\begin{equation}
	w(x) \approx \frac{1}{\sqrt{d}}\left(\pi\sqrt{V_0/E_r}\right)^{1/4} e^{-\frac{\pi^2}{2}\sqrt{V_0/E_r}(x/d)^2},
	\label{eq:wann_gauss}
\end{equation}
a valid approximation for a lattice depth of \mbox{$V_0=15 E_r$} considered throughout this work \cite{San10}. 
The form factor that follows from \eqref{eq:wann_gauss} is also Gaussian, 
\begin{align}
	W(\kappa)=e^{-\frac{(\kappa d)^2}{4\pi^2\sqrt{V_0/E_r}}}.
\end{align}
%
\section{Analysis of the inelastic cross section}
\label{sec:analysis}
%
In the following, we present a detailed study of the many-particle cross section in the Bogoliubov approximation, Eq.~\eqref{eq:BogCSfinal}. 
The condensate density is calculated for each value of $U$ from Eq.~\eqref{eq:N_depl}, and the Bogoliubov energies $\omega_q$ and the cross section are obtained accordingly. The analytical approximation is compared against the exact result of Eq.~\eqref{eq:crosssec_gen} 
from the numerical diagonalization of $H_{\rm BH}$ [Eq.~\eqref{eq:HBH}]. We emphasize that the exact calculation of the cross section requires the full spectrum of the system, which is a challenging numerical task due to the exponential growth of the underlying Hilbert space with $N$ and $L$. Thus, we restrict ourselves to small systems with a moderate number of bosons. 

\subsection{Angular dependence of the cross section}
\label{sec:angeffect}
\begin{figure}
 \includegraphics[width=\figwidth]{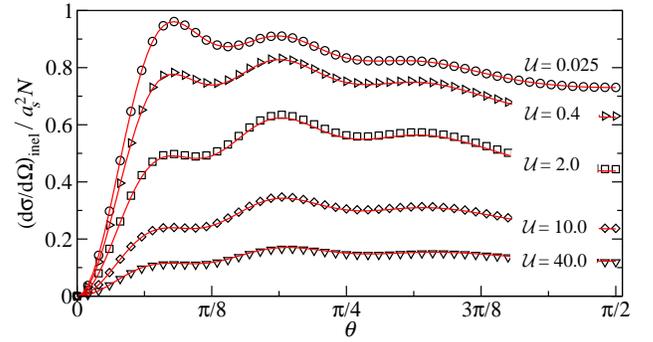}
\caption{(color online) Inelastic cross section as a function of the scattering angle $\theta$ for $N=25$ particles in $L=5$ sites, normalised by $N$, for different interaction strength $\un$. The incoming energy is 
$E_{0} = 2E_r$, the hopping energy is set to $J=0.0065 E_r$ and we use $m=M$. Symbols correspond to the exact cross section 
obtained from the diagonalisation of $H_{\rm BH}$. Red lines show the Bogoliubov approximated cross section [Eq.~\eqref{eq:BogCSfinal}]. 
The quality of the approximation is independent of the angle.}
\label{fig:CS_theta}
\end{figure}

%
The angular dependence of the cross section 
stems from the interference of terms with different phases in $\Sigma(\kappa_q-q)$ (Bragg-Scattering \cite{AshMer:SSP}), i.e.~through the dependence on the transferred momentum $\kappa_q$. For vanishing interaction, the inelastic cross section is the difference between the purely elastic Bragg signal of the Mott Insulator (Fig.~\ref{fig:CS_MISF}) and the full signal of the superfluid, leading to the structured angular dependence shown in Fig.~\ref{fig:CS_theta}.
As the interaction increases, the inelastic background decays 
and the interference features are progressively washed out.
The expression for the inelastic cross section in Bogoliubov approximation [Eq.~\eqref{eq:BogCSfinal}] provides a remarkable agreement with the exact calculations 
even for noticeably large values of $\mathcal{U}$, as can be seen in Fig.~\ref{fig:CS_theta}. We will later discuss that, in fact, the 
quality of the approximation depends non-monotonically on the interaction. 
\begin{figure}
 \includegraphics[width=.9\columnwidth]{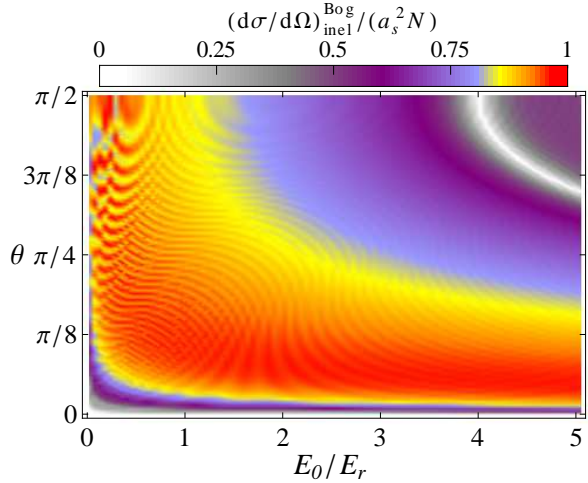}
\caption{(color online) Dependence of the inelastic Bogoliubov cross section on the incoming energy $E_0$ and the scattering angle $\theta$ for $N=100$ particles on $L=100$ sites at fixed interaction strength $U=0.02 J$, for which $\delta n/n =0.012$. We set $J=0.0065 E_r$ and $m=M$. 
}
\label{fig:CS_E0thetadep}
\end{figure}

Figure \ref{fig:CS_E0thetadep} shows the dependence of the cross section on both the scattering angle and the incoming energy of the probe. For an intermediate fixed $\theta$,  
scanning through different values of $E_{0}$ is 
similar to probing the system's spatial structure by varying the detected scattering angle.
For low $E_0$ the angular dependence of the cross section exhibits a rich oscillating pattern which changes strongly with the system size. 
In order to analyse the most prominent features of the effect of $U$ on the cross section, we will 
consider the regime where the probe energy is high as compared to the excitation spectrum, and all excited states contribute equally to the scattering signal.
This requires $E_0\gg 4J\sqrt{1+\un/2}\geqslant\omega_q$, which with our typical choice of parameters is fulfilled for $E_0\gtrsim E_r$. Let us recall that in order to avoid interband excitations the incoming energy must be smaller than the band gap of the spectrum of the lattice, $E_0<6E_r$. 
In the chosen regime for $E_0$, 
a considerable simplification of Eq.~\eqref{eq:BogCSfinal} is obtained by assuming $L\gg1$ (Appendix \ref{app:largeL}): 
\begin{equation}
    \frac{1}{Na_s^2}\bcsin= \frac{n_0}{n} \frac{\epsilon_{\kappa_{\rm el}}}{\sqrt{\epsilon_{\kappa_{\rm el}}(\epsilon_{\kappa_{\rm el}}+2Un_0)}} \left|W(\kappa_{\rm el})\right|^2,
    \label{eq:CS_contlim}
\end{equation}
where 
\begin{equation}
 \kappa_{\rm el}d=-\pi\sin\theta \sqrt{\frac{E_0}{E_r}\frac{m}{M}}
 \label{eq:kappael}
\end{equation}
is the $x$-component of the transferred momentum for elastic scattering, and the condensate density $n_0$ must still be obtained from Eq.~\eqref{eq:N_depl}. 
\begin{figure}
 \includegraphics[width=\figwidth]{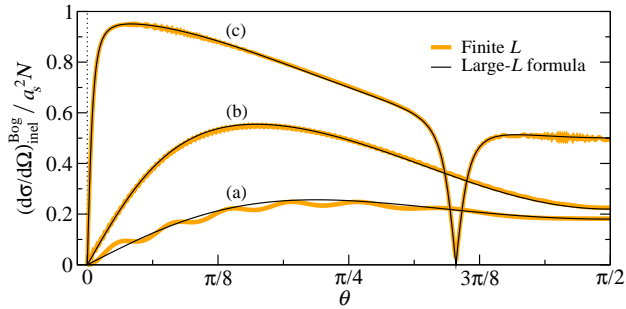}
\caption{(color online) Comparison of the large-$L$ formula of the Bogoliubov cross section [Eq.~\eqref{eq:CS_contlim}, black lines] vs expression \eqref{eq:BogCSfinal} (orange thick lines) for: (a) $L=10$, $n=10$, $E_0=2E_r$, $U=2J$ ($\delta n/n=0.080$), (b) $L=100$, $n=5$, $E_0=3E_r$, $U=0.5J$ ($\delta n/n=0.098$), and (c) $L=1000$, $n=1$, $E_0=5E_r$, $U=0.01J$ ($\delta n/n=0.027$). In all cases $J=0.0065 E_r$ and $m=M$.}
\label{fig:CS_contlim}
\end{figure}
In Fig.~\ref{fig:CS_contlim}, the Bogoliubov cross section [Eq.~\eqref{eq:BogCSfinal}] is compared to the large-$L$ simplification for different parameters. 
Expression \eqref{eq:CS_contlim} describes remarkably well the behavior of the cross section even for small system sizes, up to interference-induced oscillations which are more prominent the smaller the size, and die out as $L$ increases. 
Note that, whenever the transferred momentum equals a reciprocal lattice vector, i.e.~$\kappa_{\rm el}d=2\pi j$ for $j\in\mathbb{Z}$ ---in particular for $\theta=0$---, destructive interference makes the inelastic cross section vanish for all values of $U$.

\subsection{Dependence on the interaction}
\label{sec:Ueffect}
\begin{figure}
 \includegraphics[width=\figwidth]{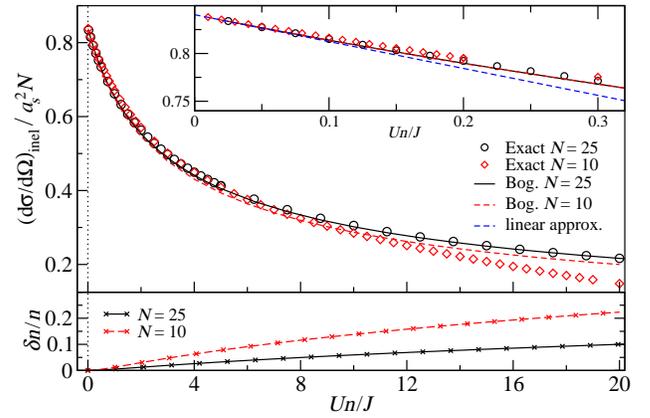}
\caption{(color online) Main panel: Inelastic cross section into the angle $\theta=\pi/4$ as a function of $\un=Un/J$ for $N = 10$ (red) and $N=25$ (black) bosons in a lattice of length $L=5$. The incoming energy is $E_{0} = 2E_r$, the hopping energy is set to $J=0.0065 E_r$ and we use $m=M$. Solid lines correspond to  
the analytical formula \eqref{eq:BogCSfinal}, while symbols are exact results. The inset shows the behavior for small values of $\un$. For $\un\ll1$, the normalized inelastic cross section 
becomes independent of the density $n$, and can be approximated by a linear function (blue, dashed) given by Eq.~\eqref{eq:lin_app1}. Lower panel: relative condensate depletion for $N=10$ (red, dashed) and $N=25$ (black, solid) particles.}
\label{fig:CS_UnJ_1}
\end{figure}

In Fig.~\ref{fig:CS_UnJ_1}, the inelastic cross section is shown as a function of $\un$ for a system of $L=5$ sites for different numbers of particles ($N=10,\,25$).
For a fixed system size and particle density $n$, and high incoming energy, the inelastic cross section decays monotonically with the interaction strength $\un$. Even for low density ($n=2$),  
expression \eqref{eq:BogCSfinal} describes the cross section correctly over a wide range of values for the interaction, 
and eventually deviates increasingly from the exact result 
as $\un$ becomes larger. 
Since the Bogoliubov treatment 
requires a high 
number of bosons, the approximation performs considerably better for $n=5$, for which we find a remarkable agreement with the exact result even for large values of $\un$. 
For the integer densities considered, a careful observation reveals that 
as $\un$ increases, one finds a regime where Eq.~\eqref{eq:BogCSfinal} underestimates slightly the cross section, until both the exact and the approximated results cross, 
and eventually ---for sufficiently large $\un$---, the Bogoliubov expression decreases slower than the exact result. 
The latter \mbox{large-$\un$} behavior can be understood qualitatively in the following way. 
For integer $n$, the condensate density $n_0$ will vanish at the SF-MI phase transition, for a finite $\un$. 
Since in the Bogoliubov approximation $n_0$ only vanishes for $U\rightarrow\infty$, a slower decay of the condensate fraction, and thus of Eq.~\eqref{eq:BogCSfinal} should be expected. 
Although the phase transition does not 
take place in a finite system, 
qualitatively, it manifests itself in the faster decay of the exact cross section for large interactions, as compared to the analytical approximation. 

\begin{figure}
\includegraphics[width=.85\columnwidth]{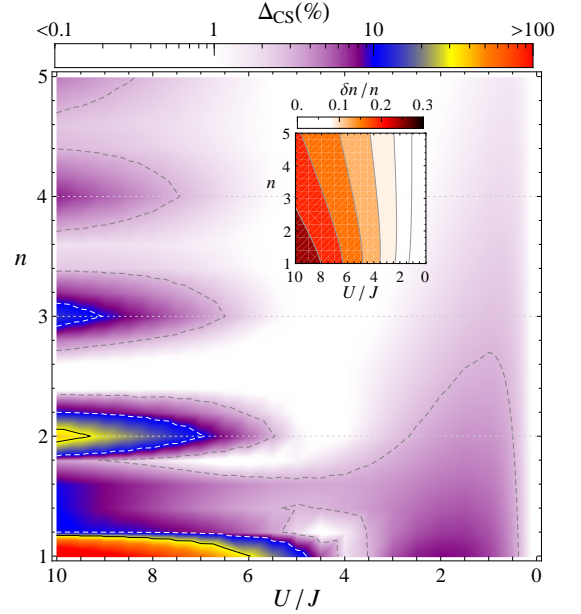}
\caption{(color online) Average relative deviation $\Delta_{\rm CS}$ of the Bogoliubov inelastic cross section with respect to the exact result, as a function of the interaction $U/J$ and the density $n$ for a system of $L=5$ sites (note the inverted $x$-axis). Relevant parameters are the same as in Fig.~\ref{fig:CS_UnJ_1}. Contour lines corresponding to $3\%$ (gray, dashed), $10\%$ (white, dashed), and $30\%$ (black, solid) deviation are marked.
The inset shows the relative condensate depletion $\delta n/n$.  
}
\label{fig:RelDiff}
\end{figure}

The above reasoning is supported 
by the behavior of the average relative deviation of Eq.~\eqref{eq:BogCSfinal}
with respect to the exact result,
\begin{equation}
	\Delta_{\rm CS} = \left\langle \frac{\left|\bcsin-\ecsin\right|}{\ecsin}\right\rangle_{[0,\pi/2]},
	\label{eq:CSdeviation}
\end{equation}
where $\langle\cdot\rangle$ 
indicates the average over the scattering angle $\theta$.  
Figure \ref{fig:RelDiff} shows $\Delta_{\rm CS}$ as a function of the density $n$ and the interaction $U$. 
A non-monotonic dependence of $\Delta_{\rm CS}$ on $U$ is observed, 
which corresponds to 
the aforementioned crossing of the analytical approximation and the exact result. 
At integer values of the density, the deviation \eqref{eq:CSdeviation} increases 
after the crossing (minimum of $\Delta_{\rm CS}$) as $U$ grows and the system approaches the Mott insulating limit.  
However, for non-integer $n$, the approximation performs always better for large $U$, as should be expected, since in this case the system remains in a SF state. 
We emphasize that this distinct behavior already manifests itself for small systems, and even for $L=5$ the analysis of $\Delta_{\rm CS}$ in Fig.~\ref{fig:RelDiff} shows qualitatively the structure of the ground-state phase diagram of $H_{\rm BH}$.
Note that, in contrast, the relative depletion 
(Fig.~\ref{fig:RelDiff}, inset) 
derived within the framework of the Bogoliubov approximation 
reveals no information on the SF-MI 
transition. 
Overall, $\Delta_{\rm CS}$ decreases with the density, showing the improving quality of the Bogoliubov approximation, reflected in the decrease of $\delta n/n$ for increasing $n$. 

\subsection{Decay for weak interaction}
\label{sec:smallU}
%
The Bogoliubov approximation, and thus Eq.~\eqref{eq:BogCSfinal}, is valid for small condensate depletion, which does not necessarily imply that the interaction parameter $\un$ is small, as demonstrated in Fig.~\ref{fig:CS_UnJ_1}. 
Since for each value of the interaction, the condensate density $n_0$ must be obtained numerically [see Eq.~\eqref{eq:N_depl}], 
there is no closed expression of Eq.~\eqref{eq:BogCSfinal} as a function of $U$. 
Nevertheless, a $\un$-expansion of the cross section keeping the full dependence of the depletion reveals that 
\begin{multline}
	\frac{1}{Na_s^2}\bcsin =  \left(1-\frac{\delta n}{n}\right)\bigg[\Gamma_{\rm sf}(L, E_{0},\theta) \\ 
	- \Lambda(L,E_{0},\theta) \left(1-\frac{\delta n}{n}\right) \un + O\left(\un^2\left(1-\frac{\delta n}{n}\right)^2\right)\bigg],
\end{multline}
where $\Gamma_{\rm sf}(L, E_{0},\theta)$ 
is the normalised superfluid inelastic cross section [cf.~Table \ref{tab:CS_MISF} and Eq.~\eqref{eq:inelasCS}], and 
\begin{align}
 	\Lambda(L,E_{0},\theta)=&\frac{J}{2L^2E_{0}}\sum_{q\neq0}\left[\frac{2E_{0}-\epsilon_q}{\epsilon_q\sqrt{1-\frac{\epsilon_q}{E_{0}}}}+\kappa_{el}\frac{\partial}{\partial\kappa_q^{\rm sf}}\right] \notag\\
	&\times\left|\Sigma(\kappa_q^{\rm sf}-q)W(\kappa_q^{\rm sf})\right|^2,
\end{align}
with $\kappa_q^{\rm sf}=\kappa_{\rm el}\sqrt{1-\epsilon_q/E_0}$ and $\kappa_{\rm el}$ given in Eq.~\eqref{eq:kappael}.

\begin{figure}
 \includegraphics[width=\figwidth]{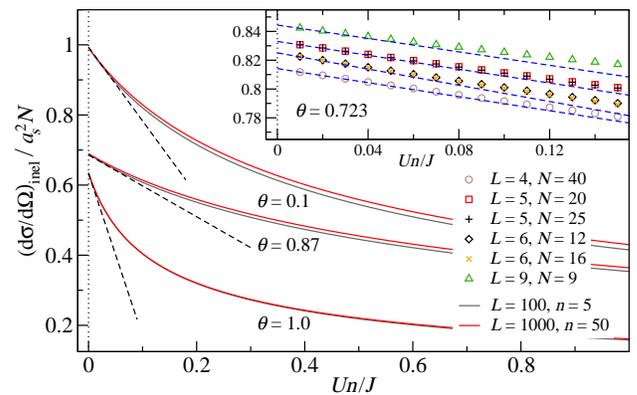}
\caption{(color online) Decay of the inelastic cross section vs $\un$, for $E_0=5E_r$, different scattering angles, system sizes and densities, as indicated in the plot. Solid lines are obtained from the Bogoliubov formula \eqref{eq:BogCSfinal} and black dashed lines depict the linear approximation given in Eq.~\eqref{eq:decaylargeL}. At $\un=1$, we have $\delta n/n=0.011$ for $L=1000$ and $n=50$, and $\delta n/n=0.054$ for $L=100$ and $n=5$. The inset shows exact numerical results (symbols) obtained for $E_0=3E_r$, and blue dashed lines correspond to the linear decay given by Eq.~\eqref{eq:lin_app1}. In all cases $J=0.0065E_r$ and $m=M$.%
}
\label{fig:lineardecay}
\end{figure}

In the regime $\un\ll1$, we know that the depletion exhibits a quadratic dependence on $\un$ [Eq.~\eqref{eq:depl_quadrappr}], and therefore for weak interaction the 
inelastic cross section behaves as 
\begin{equation}
	\frac{1}{N a_s^2}\bcsin = 
	\Gamma_{\rm sf}(L, E_{0},\theta) - \Lambda(L,E_{0},\theta)\,\un, \quad \un\ll 1.
	\label{eq:lin_app1}
\end{equation}
We thus find that the normalized inelastic cross section decays \emph{linearly} with $\un$. 
Moreover, the decay is \emph{independent} of the density $n$. 
We emphasize that, in fact, for any dependence $\delta n/n\propto\un^\mu$ with $\mu>1$, the first order 
of the cross section for non-vanishing interaction is linear and \emph{independent of the depletion}, 
and due solely to the interaction-induced change of the Bogoliubov spectrum $\omega_q$.  
The linear behavior in the emergence of the decay of the inelastic cross section ---and its independence on the density for a fixed $L$--- can be clearly observed in the inset of Figs.~\ref{fig:CS_UnJ_1} and \ref{fig:lineardecay}, where the validity of expression \eqref{eq:lin_app1} is also confirmed. 

\begin{figure}
 \includegraphics[width=\figwidth]{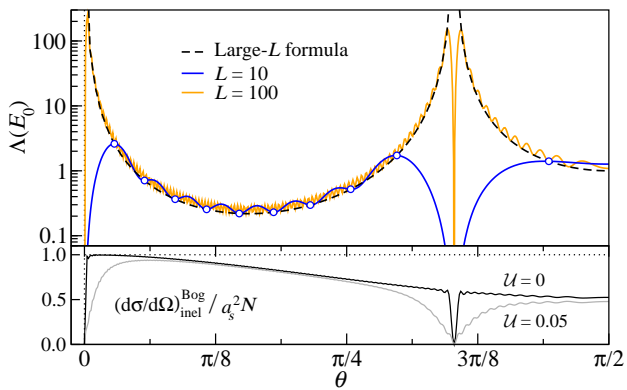}
\caption{(color online) Slope $\Lambda(L,E_0,\theta)$ of the decay of the inelastic Bogoliubov cross section at incoming energy $E_0=5E_r$, $J=0.0065E_r$ and $m=M$, for $L=10$ (blue line), $L=100$ (orange line) and the large-$L$ approximation (dashed) from Eq.~\eqref{eq:decaylargeL}. White circles highlight the angles at which the large-$L$ approximation is exact for $L=10$, given in Eq.~\eqref{eq:exactangles}. The lower panel shows the inelastic cross section for $L=100$ and $n=5$ at $U=0$ and $U=0.01J$ ($\delta n/n=5\times10^{-3}$). 
}
\label{fig:slope}
\end{figure}

As presented in Sec.~\ref{sec:angeffect} and Appendix \ref{app:largeL}, a considerable simplification of the formalism can be achieved by assuming $L\gg1$, and in the regime of high-incoming energy ($E_0\gg4J\sqrt{1+\un/2}\geqslant\omega_q$). In this case, using Eq.~\eqref{eq:CS_contlim}, the linear decay is simply given by 
\begin{equation}
 \frac{1}{N a_s^2}\bcsin = |W(\kappa_{\rm el})|^2 \left(1- \frac{\un}{4\sin^2(\kappa_{\rm el}d/2)}\right), \quad \un\ll 1, 
 \label{eq:decaylargeL}
\end{equation}
for $\kappa_{\rm el}d\neq 2\pi j$, $j\in \mathbb{Z}$. 
In Fig.~\ref{fig:slope}, we compare the slope of the decay $\Lambda(L,E_0,\theta)$ to the one given in the latter equation. 
The large-$L$ formula provides a very good approximation (up to finite-size induced oscillations) even for small systems, and for a given $L$ it is exact for certain values of the scattering angle, as discussed in Appendix \ref{app:largeL}. The expression above also reveals that it is in the vicinity of the points corresponding to $\kappa_{\rm el}d= 2\pi j$ where the decay occurs fastest (see lower panel of Fig.~\ref{fig:slope}).

For sufficiently small $\un$ the inelastic cross section must decrease linearly for any system size. One may however ask, since the $\un$-range of validity of the quadratic dependence of the depletion decreases with $L$, whether the linear decay plays a relevant role for large system sizes. Indeed, as we demonstrate in Fig.~\ref{fig:lineardecay}, if the atomic density is high enough, the linear approximation given in Eq.~\eqref{eq:decaylargeL} describes correctly the behavior of the decay as $\un\rightarrow0$, even for large system sizes. Furthermore, we emphasize that in this case the decay is not only independent of the density, but also of the system size, and determined uniquely by the incoming energy and the scattering angle. 

%
\section{Conclusion}
\label{sec-concl}
In this work 
we have studied the inelastic cross section of a coherent matter-wave 
scattered from 
interacting ultracold bosons in an optical lattice, focusing on 
the dependence on the interaction among the trapped bosons. 
For this purpose, we have used Bogoliubov's formalism to obtain analytically the inelastic cross section in the regime of 
small condensate depletion. We have compared the analytical results against exact numerical calculations, and we have analysed the cross section with respect to 
the scattering angle, incoming energy, density, system size and interaction energy $U$.
We found a linear decay of the normalized cross section for weak interaction.  
For a given incoming energy and scattering angle, the decay with the interaction strength $\un$ 
is independent of the number of particles in the system, of the condensate depletion, and (above a certain number of sites) becomes also independent of the system size. 

Here we considered mainly the regime of high incoming probe energy, where all the excitation spectrum of the system contributes comparably to the scattering signal. 
The analysis for low incoming energy, as well as the energy-resolved scattering, would additionally provide access to the spectral information of the target \cite{HunHC12}.

While in the Mott insulating limit ($U\rightarrow\infty$) the inelastic cross section vanishes \cite{SanMH10}, whether it can be used to characterize the SF-MI transition  remains to be demonstrated. For instance, it is not known if the vanishing of the inelastic cross section extends to the whole Mott lobes, or if on the contrary the transition might be signalled by a change in the dependence of the decay on $U$ as the phase boundary is crossed. Since the transition cannot be reached within the Bogoliubov prescription, different techniques will have to be used to answer this question. 

\begin{acknowledgments}%
 We gratefully acknowledge the Deutsche Forschungsgemeinschaft for financial support.
 We thank C.~A.~M\"uller for insightful discussions, and V.~Shatokhin for useful comments and 
 the careful reading of the manuscript. 
\end{acknowledgments}
\appendix
\section{Cross section for large system size}
\label{app:largeL}
\begin{figure}
 \includegraphics[width=.95\columnwidth]{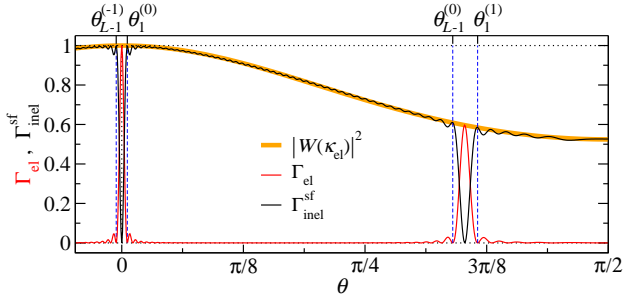}
 \caption{Elastic cross section [Eq.~\eqref{eq:elasCS}, red] and inelastic cross section in the SF limit [Eq.~\eqref{eq:inelasCS}, black] for a lattice of size $L=50$ with $E_0=5E_r$, $V_0=15E_r$, $J=0.0065E_r$ and $m=M$. The thick orange line shows the form factor $|W(\kappa_{\rm el})|^2$, which determines both cross sections for large $L$ according to Eqs.~\eqref{eq:elasCSlargeL} and \eqref{eq:inelasCSlargeL}. Note the different normalization factors: in a system with $N$ bosons, the elastic cross section in units of $a_s^2$ is $N^2\Gamma_{\rm el}$ and the inelastic $N\Gamma_{\rm inel}^{\rm sf}$. Vertical dashed lines mark the intervals given in Eq.~\eqref{eq:thetainterval} .}
 \label{fig:sfCSapp}
\end{figure}
In the non-interacting case ($U=0$), i.e.~in the SF limit, the elastic and inelastic cross sections converge to well-defined expressions as $L\rightarrow\infty$. 
As given in Table \ref{tab:CS_MISF}, the elastic cross section (neglecting off-diagonal overlapping of the Wannier functions) reads
\begin{equation}
 \Gamma_{\rm el}\equiv \frac{1}{N^2 a_s^2}\csel= \frac{1}{L^2}|\Sigma(\kappa_{\rm el})|^2 |W(\kappa_{\rm el})|^2, 
 \label{eq:elasCS}
\end{equation}
in terms of $\Sigma(\kappa)$ and $\kappa_{\rm el}$ defined in Eqs.~\eqref{eq:sigma} and \eqref{eq:kappael}, respectively. 
In the large-$L$ limit we observe that 
\begin{equation}
 \frac{1}{L^2}|\Sigma(\kappa_{\rm el})|^2 \;\underset{L\rightarrow\infty}{\longrightarrow}\; \delta_{\kappa_{\rm el},Q},
\end{equation}
for all reciprocal lattice vectors $Q=2\pi j /d$, $j\in\mathbb{Z}$. Thus, 
\begin{equation}
 \Gamma_{\rm el} = |W(\kappa_{\rm el})|^2 \delta_{\kappa_{\rm el},Q}, \quad L\gg 1.
 \label{eq:elasCSlargeL}
\end{equation}
On the other hand, the normalized inelastic cross section is given by 
\begin{align}
 \Gamma_{\rm inel}^{\rm sf} &\equiv\frac{1}{N a_s^2}\csin \notag\\ &= \frac{1}{L^2}\sum_{q\neq 0} 
 \sqrt{1-\frac{\epsilon_q}{E_0}}|\Sigma(\kappa_q^{\rm sf}-q) W(\kappa_q^{\rm sf})|^2, 
 \label{eq:inelasCS}
\end{align}
where $\kappa_q^{\rm sf}=\kappa_{\rm el}\sqrt{1-\epsilon_q/E_0}$, and $\epsilon_q$ is the 1-D Bloch dispersion relation [Eq.~\eqref{eq:lattdisp}].
For large $L$ one has
\begin{align}
 \frac{1}{L}|\Sigma(\kappa_q^{\rm sf}-q)|^2 \;& \underset{L\rightarrow\infty}{\longrightarrow} \;\frac{2\pi}{d}\delta\left(\kappa_q^{\rm sf}-q-Q\right), \label{eq:deltalargeL}\\
 \frac{1}{L}\sum_{q\neq0} \; & \underset{L\rightarrow\infty}{\longrightarrow} \; \frac{d}{2\pi} \int dq, 
 \label{eq:intlargeL}
\end{align}
where the $q$-sum in the first Brillouin zone, is replaced by an integral over the interval $(0,2\pi/d)$ excluding the borders. One can then write 
\begin{equation}
 \Gamma_{\rm inel}^{\rm sf}=\int dq\,\delta(\kappa_q^{\rm sf}-q)\sqrt{1-\frac{\epsilon_q}{E_0}}|W(\kappa_q^{\rm sf})|^2,
\end{equation}
after making the change $q+Q\rightarrow q$. The integration interval now runs over all space excluding the points $q=Q$.
The latter expression evaluates to 
\begin{equation}
 \Gamma_{\rm inel}^{\rm sf}=\sqrt{1-\frac{\epsilon_{q'}}{E_0}} \frac{|W(q')|^2}{\left|1+ \kappa_{\rm el}d\frac{J\sin(q'd)}{E_0\sqrt{1-\epsilon_{q'}/E_0}} \right|},
\end{equation}
where $q'\neq Q$ is the solution of $\kappa_{q'}^{\rm sf}-q'=0$ (for our choice of $J=0.0065E_r$ one can see that there is only one possible $q'$). 
If $q'=Q$ then $\Gamma_{\rm inel}^{\rm sf}=0$. In the case of high incoming probe energy, $E_0\gg 4J$, the expression above can be further simplified and the cross section converges to  
\begin{equation}
  \Gamma_{\rm inel}^{\rm sf}= |W(\kappa_{\rm el})|^2 (1-\delta_{\kappa_{\rm el},Q}), \quad L\gg1,\,E_0\gg 4J.
\label{eq:inelasCSlargeL}
\end{equation}
Figure \ref{fig:sfCSapp} shows $\Gamma_{\rm el}$ and $\Gamma_{\rm inel}^{\rm sf}$ for a lattice with $L=50$ sites and how they compare with the large-$L$ limit.

Similarly, a simpler expression of the inelastic cross section in Bogoliubov approximation [Eq.~\eqref{eq:BogCSfinal}] can be obtained after the assumption $L\gg1$. 
From Eqs.~\eqref{eq:deltalargeL} and \eqref{eq:intlargeL}, and following the same procedure as for $\Gamma_{\rm inel}^{\rm sf}$, we obtain 
\begin{equation}
\frac{1}{Na_s^2}\bcsin = \frac{n_0}{n} \sqrt{1-\frac{\omega_{\tilde q}}{E_0}} \frac{\epsilon_{\tilde q}}{\omega_{\tilde q}}
\frac{\left|W(\tilde q)\right|^2}{\left| 1+ \kappa_{\rm el}d\frac{J\sin(\tilde qd)(\epsilon_{\tilde q} +U n_0)}{E_0\omega_{\tilde q}\sqrt{1-\omega_{\tilde q}/E_0}}\right|},
\end{equation}
where $\tilde q$ is the solution of $\kappa_{\tilde q}-\tilde q=0$ (for our typical choice of $U$ and $J=0.0065E_r$ there is only one possible $\tilde q$), and let us recall that $\kappa_q=\kappa_{\rm el}\sqrt{1-\omega_q /E_0}$. 
In the regime of high incoming probe energy, $E_0\gg 4J\sqrt{1+\un/2}\geqslant\omega_q$, the expression above is well approximated by 
\begin{equation}
 \frac{1}{Na_s^2}\bcsin = \frac{n_0}{n} \frac{\epsilon_{\kappa_{\rm el}}}{\omega_{\kappa_{\rm el}}}\left|W(\kappa_{\rm el})\right|^2,
 \label{eq:CS_contlimapp}
\end{equation}
which corresponds to Eq.~\eqref{eq:CS_contlim}. Note that an $L$-dependence still exists through the value of the condensate fraction $n_0$, which is given by Eq.~\eqref{eq:N_depl}. 
This large-$L$ expression describes also remarkably well the qualitative behavior of the cross section for systems as small as a $10$-site lattice, as shown in Fig.~\ref{fig:CS_contlim}. Since Eq.~\eqref{eq:CS_contlimapp} relies on the identification $q=\kappa_{\rm el}$, the limitation of the applicability of the approximation for a finite $L$ stems from the discretized nature of the contributing quasimomentum, $q\in[2\pi/Ld,2\pi(L-1)/Ld]+Q$ with $\Delta q =2\pi/Ld$.
This leads to a range for the scattering angle within which Eq.~\eqref{eq:CS_contlimapp}, and equivalently Eqs.~\eqref{eq:elasCSlargeL} and \eqref{eq:inelasCSlargeL}, can be expected to perform well for finite systems (cf.~Fig.~\ref{fig:sfCSapp}), namely  
\begin{equation}
 \left[\theta_1^{(j)},\theta_{L-1}^{(j)}\right],\qquad j\in \mathbb{Z},
 \label{eq:thetainterval}
\end{equation}
where 
\begin{equation}
 \theta_s^{(j)}=\arcsin\left(2\sqrt{\frac{E_r}{E_0}\frac{M}{m}}\left[j+ \frac{s}{L}\right]\right), \quad s=1,\ldots,L-1.
 \label{eq:exactangles}
\end{equation}
In fact, one can see that in the regime of high incoming probe energy, for a finite system of size $L$, the large-$L$ expressions are \emph{exact} at the scattering angles $\theta^{(j)}_s$.


\end{document}